\documentclass[traditabstract]{aa} % for the abstract without structuration
\usepackage{graphicx}
\usepackage{txfonts}
\usepackage{color}
\usepackage{natbib}
\usepackage{verbatim}

\begin{document}
\newcommand{\um}{\,$\mu$m\,}
\newcommand{\hii}{H\,{\sc ii}\,}
\newcommand{\cv}{C$_{24}$\,} 
\newcommand{\cs}{C$_{60}$\,}
\newcommand{\csp}{C$_{60}^+$}
\title{Top-down formation of fullerenes in the interstellar medium} %:\\ proof of concept by photochemical modeling}
\author{O. Bern\'e \inst{1,2}, J. Montillaud \inst{3,4} and C. Joblin \inst{1,2}}
\institute{Universit\'e de Toulouse; UPS-OMP; IRAP;  Toulouse, France
\and CNRS; IRAP; 9 Av. colonel Roche, BP 44346, F-31028 Toulouse cedex 4, France
\and Department of Physics, PO Box 64, University of Helsinki, 00014, Helsinki, Finland
\and Institut Utinam, CNRS UMR 6213, OSU THETA, Universit\'e de Franche-Comt\'e, 41bis avenue de l'Observatoire, 25000 Besan\c{c}on, France}
%\author{Julien Montillaud}
%\institute{Department of Physics, PO Box 64, University of Helsinki, 00014, Helsinki, Finland}
%\author{Christine Joblin}
\titlerunning{Formation of fullerenes in the ISM}
\authorrunning{}
\date{Received August ??, 2012; accepted ??, 2012}

\abstract{Fullerenes have been recently detected in various circumstellar and interstellar environments,
raising the question of their formation pathway.
It has been proposed that they can form at the low densities found in the interstellar medium by the photo-chemical
processing of large polycyclic aromatic hydrocarbons (PAHs).
Following our previous work on the evolution of PAHs in the NGC 7023 reflection nebula, we evaluate,
using photochemical modeling, the possibility that {the PAH} C$_{66}$H$_{20}$ (i.e. circumovalene) can 
lead to the formation of \cs upon irradiation by ultraviolet photons.
%In this letter, we evaluate this proposal, using a detailed photochemical model with an explicit description of the 
%internal energy of the molecules, and using state of the art values for the dissociation parameters for every reactions.
%We follow the evolution of {the PAH} C$_{66}$H$_{20}$ (circumovalene) for physical conditions as those found in the NGC 7023 
%reflection nebula, where it has been shown that  C$_{60}$ is formed. 
The chemical pathway involves full dehydrogenation of C$_{66}$H$_{20}$, folding into a floppy
closed cage and shrinking of the cage by loss of C$_2$ units until it reaches the symmetric C$_{60}$ molecule.
At 10'' from the illuminating star and with realistic molecular parameters, the model predicts that 100\% 
of C$_{66}$H$_{20}$ is converted into C$_{60}$ in $\sim$ 10$^5$ years, a timescale comparable 
to the age of the nebula.  {  Shrinking appears to be the kinetically limiting step of the whole process. 
Hence, PAHs larger than C$_{66}$H$_{20}$  are unlikely to contribute significantly to the formation of C$_{60}$, 
while PAHs containing between 60 and 66 C atoms should contribute to the formation of C$_{60}$ 
with shorter timescales, and PAHs containing less than 60 C atoms will be destroyed.} 
Assuming a classical size distribution for the PAH precursors, our model predicts absolute abundances of 
\cs are up to several $10^{-4}$ of the elemental carbon, {  i.e. less than a percent of the typical interstellar PAH 
abundance}, which is consistent with observational studies.
According to our model, once formed, C$_{60}$ can survive much longer 
($>10^7$ years for radiation fields below $G_0=10^4$) than other fullerenes because of the
remarkable stability of the \cs molecule at high internal energies.
Hence, a natural consequence
% of the top-down photo-chemistry described in this paper 
is that \cs is more abundant than other fullerenes in highly irradiated environments.
} 

\maketitle

\section{Introduction}

The mid-infrared (mid-IR) spectrum of galactic and extragalactic objects exhibits emission in bands (strongest at 3.3, 6.2, 7.7, 8.6, and 11.2 $\mu$m)
attributed to carbonaceous macromolecules, more specifically to polycyclic aromatic hydrocarbons (PAHs,  \citealt{tie08}). In addition to PAH bands, IR emission bands at 7.0, 8.5, 17.4, and 19.0 $\mu$m have been reported \citep{cam10,sel10}, and found to 
match quite closely the IR active bands of neutral buckminsterfullerene (C$_{60}$,~\citealt{kro85}), a cage-like carbon molecule.
Additional bands at 6.4, 7.1, 8.2, and 10.5 $\mu$m were observed in the NGC 7023 reflection nebula and attributed to 
the C$_{60}^+$ cation \citep{ber13}. Carbonaceous macromolecules,  especially  PAHs, are believed to play 
a fundamental role in the physics and chemistry of the interstellar medium ({ISM)}, and their infrared signatures are commonly 
used as a tracer of physical conditions (especially the UV radiation field). 
PAHs are believed to lock up between 5 and 20$\%$ of the elemental carbon  \citep{job11}, while
\cs and \csp are found in small abundances in the ISM (at most $5.6\times10^{-4}$ 
of the elemental carbon for the neutral form according to \citealt{cas14}, and at most $10^{-4}$ of the elemental carbon for the cation
according to \citealt{ber13}). Nevertheless, \cs and \csp\, are the only species belonging to the family of carbonaceous macromolecules
which have been specifically identified in the ISM, and therefore these molecules have attracted considerable interest
because they open a new possibility to probe carbon chemistry and physical conditions in the ISM.

One question related to fullerenes, and in particular \cs, concerns their formation pathway.
Recently, ``top-down" schemes where larger carbon clusters shrink to reach \cs  have been proposed \citep{chu10, zha13, pie14},
and can be opposed to the traditional ``bottom-up" approach where \cs is built up from small compounds \citep{kro88, hea92, hun94, dun13}.
Using infrared observations of the NGC 7023 nebula, \citet{ber12} found evidence of an increase of the abundance 
of \cs with increasing UV field, while the abundance of PAHs decreases. This was interpreted by these authors as 
evidence for the formation of \cs from large PAHs ($N_C>60$)  under UV irradiation, a top down mechanism similar
to the one observed by  \citet{chu10}. \citet{her10, her11} and \citet{mic12} proposed a similar mechanism where the starting 
materials are more complex, such as hydrogenated amorphous carbon instead of PAHs. Top-down scenarios are 
particularly appealing, given that the densities prevailing in the ISM are many orders of magnitude too low to allow 
for a ``bottom-up" formation (i.e. starting from small compounds) over reasonable timescales\footnote{While writing the present paper, it was proposed by \citet{pat14} that nucleation of 
C atoms leading to the formation of cages can operate in the ISM. However, the densities and relative abundances of 
C and H adopted in this latter work are far from those considered here and actually observed in the interstellar medium.}. 
In this paper, we use a detailed photochemical model coupled to a description 
of the physical conditions in the NGC 7023 nebula. We demonstrate that the formation of \cs by UV processing of 
large PAH molecules is a plausible mechanism to account for astronomical observations.

\section{Model for the formation of \cs from PAHs}

\subsection{Description of the proposed scenario}
\begin{figure*}
\begin{center}
\includegraphics[width=5.5cm, angle=90]{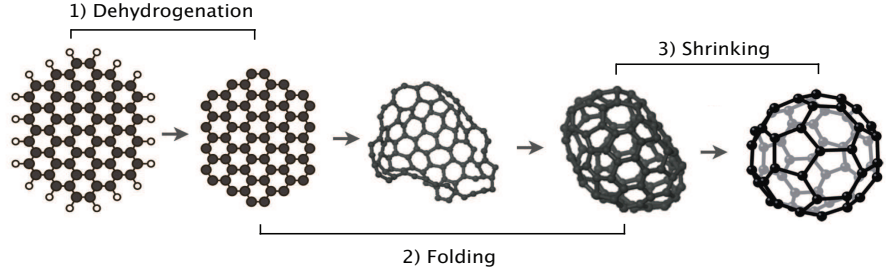}
\caption{Schematic representation of the evolutionary scenario for the formation of fullerenes from PAHs under UV irradiation.
 \label{fig_scenario}}
\end{center}
\end{figure*}

The proposed scenario of formation is inspired by the one described in \citet{ber12} and is represented 
schematically in Fig.~\ref{fig_scenario}. PAHs are assumed to be formed in the envelopes of evolved stars \citep{fre89, che92, mer14},
and then injected in the ISM. Under UV irradiation, large PAHs, (N$_C>60$) are first fully dehydrogenated 
into small graphene flakes, dehydrogenation being by far the dominant dissociation channel \citep[see][and references 
therein]{mon13}. 
Further UV irradiation { enables} these flakes to fold into closed cages. Once the cages are closed, they can loose
C$_2$ units if they continue absorbing energy \citep{irl06}.  Because of the low densities prevailing in the considered
regions ($n_H<10^4$cm$^{-3}$), the reverse reaction, i.e. { addition} of C$_2$ is { too slow to balance 
photodissociation}. Once a system has reached C$_{60}$ it will remain in this form for a very long period of time,
because of its remarkable stability.

%\vspace{-0.5cm}

\subsection{The specific case of the transformation of C$_{66}$H$_{20}$ into C$_{60}$}

In this paper we consider the evolution of { circumovalene, C$_{66}$H$_{20}$ (first molecule in Fig.~\ref
{fig_scenario})}, for the physical conditions of NGC 7023 (see Sect.~\ref
{sec:physical_conditions}). We restrict ourselves to a single molecule as a starting point, { for simplicity, and because modeling a 
complete population would introduce many free parameters that would be irrelevant in this proof of concept study. 
 C$_{66}$H$_{20}$ is selected as a test case because it gathers the typical properties expected for an interstellar PAH
($N_{\rm C}\gtrsim50$, compactness), it can be considered as a potential precursor of C$_{60}$ and its spectroscopic properties
are available in the Cagliari Theoretical spectral database of PAHs\footnote{astrochemistry.ca.astro.it/database/}
(\citealt{mal07}, hereafter Cagliari database).}
The photophysical modeling relies on several key processes: 1) UV absorption which brings the molecules to high
temperatures, 2) radiative cooling, 3) unimolecular dissociation, 4) isomerization and more specifically folding of the
graphene flakes and 5) reactions with electrons, H atoms or C$^+$. In the following sections, we present the 
adopted methods to describe these processes which were then implemented in the photochemical model developed 
by \citet{mon13}.

\subsubsection{UV absorption}

UV absorption is the first step in the photophysical evolution of the considered species. The efficiency of this process
in a given radiation field depends on the UV absorption cross-section. 
The UV absorption cross-section of C$_{66}$H$_{20}$ used here is taken from the Cagliari database.
The same value is used for dehydrogenated PAHs C$_{66}$H$_{n}$ as well as for planar C$_{66}$, considering that the UV absorption cross-section
in a PAH is, in first approximation, proportional to the number of carbon atoms. For \cs, we adopt the UV absorption cross-section from \citet{ber99}. 
The same absorption cross-section is used for the other cages, but with a scaling proportional to the number of C atoms \citep{job92}.
%The ionization cross section for C$_{66}$H$_{20}$ is as in \citet{mon13}. For the pure carbon clusters, we only consider neutral 
%species since observations indicate that neutral fullerenes dominate over the cations (except in regions of very high 
%radiation fields, see \citealt{ber13}). 
The impact of the chosen UV absorption cross-sections on the results is discussed in Sect.~\ref{sect_sensitivity}.

\subsubsection{Ionization}

In the frame of our interstellar model, two ionization processes should be considered: direct photoionization and thermionic emission.
Direct ionization occurs after the absorption of a single UV photon with an energy above the ionization threshold.
In the case of thermoionic emission, the successive absorption of several UV photons brings the molecule to high
temperatures and can lead to delayed electron emission.

For the hydrogenated and dehydrogenated PAHs, the direct photoionization process is included, following \citet{mon13}.
For the cages, including \cs, we do not include direct photoionization. This assumption relies on the fact that observations 
indicate that this species is mostly in the neutral form in NGC 7023, \csp being localized only very close to the star \citep{ber13}. 
It cannot be excluded that this is due to an efficient recombination of \cs with electrons from the ambient medium, and hence ionization could still be efficient as 
a sink of energy for relaxation. Since the ionization potential of \cs is 7.5 eV, only photons with energies 
$7.5<h\nu<13.6$ eV can contribute to ionization. Given the spectral profile of the radiation field in NGC 7023
(see \citealt{mon13}) and the fact that in this range the ionization yield is low compared to 
the total photo-absorption \citep{ber99}, only a small fraction of the energy from absorbed UV photons will be lost 
in photoelectrons, and most of it will be converted into vibrational energy that can eventually be used for dissociation.
The other cages have similar ionization potentials \citep{sei96} and are therefore assumed to behave like \cs.
Hence, we have neglected direct photoionizaiton for the cages in the photo-chemical model. 

Thermoionic emission becomes efficient only at internal energies higher than $\sim$ 30 eV \citep{han03}. It could therefore
be an efficient relaxation process, competing with photodissociation (Sect.~\ref{sec:phd}) and cooling by fluorescence
from thermally excited electronic states (Sect.~\ref{sec:cool}). 
As can be seen in Fig.~\ref{fig_cooling}, the termoionic emission rate becomes larger than fluorescence
from thermally excited electronic states only for internal energies above $\sim$ 45 eV. This is
well above the internal energies of dissociation (see Sect.~\ref{internal_energies}) and therefore relaxation through 
thermionic emission can be neglected since it will always be dominated by fluorescence
from thermally excited electronic states.

\subsubsection{Radiative cooling}
\label{sec:cool}

Radiative cooling can occur through emission of photons in the infrared, but when the molecules reach high 
enough temperatures, it has been shown that fluorescence
from thermally excited electronic states becomes the dominant cooling mechanism
(see \citealt{leg88}, who called this mechanism Poincar\'e fluorescence).
 This process has been observed for internal energies of $\sim7$ eV for the anthracene cation, C$_{14}$H$_{10}^{+}$,
 \citep{mar13} and internal energies of $\sim 13$ eV for the \cs anion \citep{and01}.
  
\begin{figure}
\begin{center}
\includegraphics[width=\hsize]{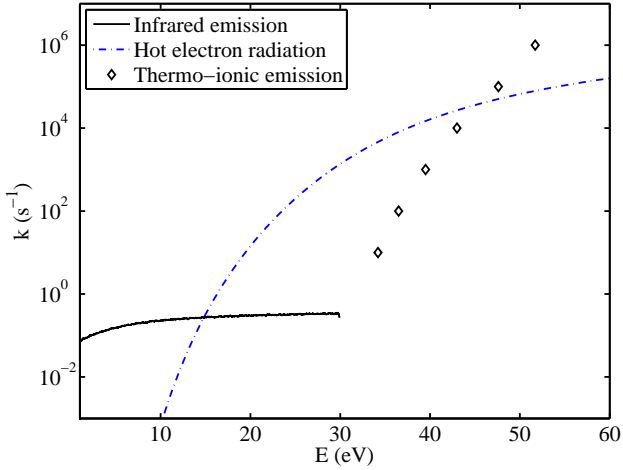}
\caption{Cooling rates for \cs as a function of the internal energy of the molecule. 
The infrared emission and cooling by fluorescence
from thermally excited electronic  states are calculated (labelled Poincar\'e in the figure) 
following the formalism described in this paper. The thermionic emission rates are taken from \citet{han03}. 
\label{fig_cooling}}
\end{center}
\end{figure}

The infrared {cooling} rates are calculated using the microcanonical formalism of
 \citet{job02}. This requires the full knowledge of the vibrational properties of the considered species. 
 For C$_{66}$H$_{20}$, we use the spectroscopic properties from the Cagliari database as in \citet{mon13}.
 For \cs, they are  taken from \citet{sch01}. For dehydrogenated PAHs C$_{66}$H$_{n}$ and cages, we use the DFT vibrational 
 frequencies of C$_{66}$H$_{20}$ from the Cagliari database, after removing the C-H vibrational modes and 3 modes per missing C atom.
The cooling rates calculated with this approach for \cs are reported in Fig.~\ref{fig_cooling}.

Cooling by fluorescence from thermally excited electronic states is calculated using the formalism described in \citet{chu97}. The cooling rate is given by:
 \begin{equation}
 k(T)=(2c/a_0)\sum f_i(h\nu_i/mc^2)^2~exp(-h\nu_i/k_BT) \label{eq_chupka}
 \end{equation}
 where $a_0$ is the Bohr radius, $f_i$ and $\nu_i$ are respectively the oscillator strengths and frequencies of the electronic transitions
 and $T$ is the temperature of the molecule. For C$_{66}$H$_{20}$ and for the dehydrogenated PAHs we use the $\nu_i$ and $f_i$
 values for C$_{66}$H$_{20}$ from the Cagliari database (Table~\ref{tab_electronic}). 
 For the cages, we use the energies and oscillator strengths of \cs taken from \citealt{chu97}, reproduced in Table~\ref{tab_electronic}.
 {   Note that these oscillator strengths are a convenient approximation 
 to calculate the radiative cooling but are not true oscillator strengths. Nevertheless, \citet{chu97} have shown that  
 this approximation is in very good agreement with rate calculations including detailed molecular  
 property information.} Using the microcanonical formalism described above for the calculation of the IR cooling rates, we derived the relation between 
 internal energy $E$ and temperature $T$ for the different species. Using Eq.~\ref{eq_chupka}, this allowed us to derive the 
 radiative cooling rates as a function of $E$. For the case of \cs these rates are shown in Fig.~\ref{fig_cooling}) 
 and are in good agreement with the earlier work models of \citet{tom03}. 
% \citet{and01} quote a value of $3.3\times10^4$ eV/s for \cs anion at 3000K, also in good agreement with our results. 

\begin{table}

\caption{Frequencies and oscillator strengths adopted in the approximation 
to calculate the cooling by fluorescence from thermally excited electronic states (see text for details). }
\label{tab_electronic}
\begin{center}
\begin{tabular}{ccc}
\hline\hline
$\lambda$ ($\AA$) & $\nu_i$ (Hz) & $f_i$\\
\hline
\multicolumn{3}{c}{C$_{66}$H$_{20}$}\\
\hline
5470 &$5.48\times10^{14}$ & 0.72\\
4980 &$6.02\times10^{14}$ & 0.24\\
\hline
\multicolumn{3}{c}{\cs}\\
\hline
4540 &$6.60\times10^{14}$ & 0.214 \\
3485 &$8.60\times10^{14}$ & 1.17 \\
2725&$1.10\times10^{15}$ & 2.02 \\
2240&$1.34\times10^{15}$ & 1.30 \\

\hline

\\

\hline
\end{tabular}
\end{center}
\end{table}

\subsubsection{Photodissociation: dehydrogenation and shrinking by loss of C$_2$}
\label{sec:phd}

Photodissociation is treated using the statistical approach, and based on the
inverse Laplace transform of the Arrhenius equation (see details in \citealt{mon13} and references therein).
The parameters involved in this calculation are the activation energy $E_{0}$, the pre-exponential factor $A$, { and 
the vibrational densities of states that are computed using the vibrational frequencies}. For PAH 
dehydrogenation (first step in Fig.~\ref{fig_scenario}), we use the parameters given in \citet{mon13}. 
{ For the shrinking of cages by loss of C$_2$ molecules (step 3 in Fig.~\ref{fig_scenario}), experimental and theoretical results are scarce, 
except in the case of \cs and \cs$^+$ for which published molecular data have a large scatter, as noticed by \citet{mat01}. 
These authors reanalyzed published experiments of \cs$^+$ dissociation and showed that when using similar and consistent sets 
of molecular data, most experiments lead to similar results, with a pre-exponential factor of $5 \times 10^{19}$s$^{-1}$ 
and a dissociation energy of $10.0 \pm 0.2$ eV. The activation energy should be equal to the dissociation energy if there 
is no barrier for the reverse reaction, i.e. C$_2$ addition. Experimental \citep{las97} and theoretical \citep{yi93} 
studies indeed indicate the absence of such a barrier. More recent studies are generally in line with these results 
\citep[among others]{dia03, glu04}. 
We consider each shrinking step from C$_{66}$ to C$_{58}$ individually (Eqs. 2 to 6 in Table~\ref{tab_act}).  Dissociation energies
have been determined theoretically \citep{zha92} and the values differ significantly from experimental data \citep{glu04}.
The theoretical values were computed directly for neutral cages in a consistent manner for all species. 
In contrast, the experimental values were derived from measurements on fullerene cations. 
We used the theoretical values in our standard model (Table~\ref{tab_act}).
This choice has the advantage of providing consistent energies for all cages in this study, but with a relatively high
value for \cs (11.1 eV), compared to the results of \citet{mat01} (10.0 eV). To reconcile our choice of binding energies 
with the results of \citet{mat01}, we adopted a pre-exponential factor of $2 \times 10^{20}$s$^{-1}$ which leads to a 
dissociation rate comparable to that of \citet{mat99} (their Fig.~8). We used the same factor for the C$_2$ loss of all 
cages. The impact of all these assumptions is discussed in Sect.~\ref{sect_sensitivity}.
}

\subsubsection{Folding}

For the folding of the carbon flakes (step 2 in Fig.~\ref{fig_scenario}), there are, to our knowledge, no experimental data to constrain the 
activation energy and pre-exponential factors. Recently, \citet{leb12} conducted molecular dynamics simulations of the 
folding of the C$_{96}$ and C$_{384}$ graphene nanoflakes. {   For these two systems, they ran simulations at several 
temperatures and found that the rates follow an Arrhenius law from which they were able to derive the effective 
activation energies and pre-exponential factors.} They show that these 
parameters do not vary significantly with size\footnote{It should be noted that this assertion concerns the 100-400 C atoms
range and may not extend to lower C numbers, yet as discussed in Sect.~\ref{sect_sensitivity} the activation energy would need to be
different from the adopted one by a large factor to alter our results}, hence we adopt their values reported for C$_{96}$ to describe the folding (see 
Table~\ref{tab_act}). 

\subsubsection{Reaction with electrons, H atoms or C$^+$}

The other processes of interest include the recombination with electrons and the reaction with H atoms. During the dehydrogenation step
all these processes are included, following \citet{mon13}. 
{We do not include the addition of C$_2$, because the abundance of this species is expected to be very low in the 
cavity of NGC 7023. The C$^+$ cation is however abundant, and could react efficiently with cages. 
There are to our knowledge no studies of the kinetics of this reaction. We estimated
the maximum contribution of this process in the case of \cs using the Langevin rate. A polarizability of $\sim$ 80 \AA$^3$\, for 
\cs \citep{zop08} leads to $k_{\rm Langevin} \sim 6\times10^{-9}$ cm$^{3}$s$^{-1}$, and assuming a high abundance of C$^{+}$ of 
$3\times10^{-4} \, n_{\rm H}$, one gets an effective C$^+$ addition rate of $\sim 2\times10^{-12} \, n_{\rm H}$ s$^{-1}$
(that is $< 4\times10^{-8}$ s$^{-1}$ for $n_H<2\times10^4$cm$^{-3}$). 
This rate must be compared to the rate of the reverse process, i.e. photodissociation of C$_{61}^{+}$. We modeled this
latter process with the same formalism as described in Sect.~\ref{sec:physical_conditions}, using a dissociation energy of $4$ eV \citep{sla94}
and a pre-exponential factor of $1.6\times10^{15}$ s$^{-1}$ \citep{klo91}. We find photodissociation rates from $\sim7\times10^{-8}$ 
s$^{-1}$ to $\sim2\times10^{-5}$ s$^{-1}$ for the various astrophysical conditions considered in Sect.~\ref{abundances}.
Therefore, the photodissociation of C$_{61}$ is estimated to be generally much faster than C$^{+}$ addition in the regions of interest, 
and we do not include this latter process in our study.}

\begin{table}
\caption{Activation energy and pre-exponential factors for the key reactions considered in the model}
\label{tab_act}
\begin{center}
\begin{tabular}{lcccc}
\hline \hline
 & 				& Act. energy & Pre-exp. factor	\\
%\hline
&				 & $E_a$ (eV) & $A$  (s$^{-1}$) 	 \\	
\hline
\hline
\vspace{0.2cm}
{ (1)}&{C}$_{66}^{planar} \rightarrow~$C$_{66}^{cage}$ & 4.0  & $1\times10^{15}$\\
\vspace{0.2cm}
{ (2)}&{C}$_{66}^{cage} \rightarrow~$C$_{64}^{cage}$~+~C$_2$ & 8.1$^a$ / 9.4$^b$ & $2\times10^{20}$\\
\vspace{0.2cm}
{ (3)}&{C}$_{64}^{cage} \rightarrow~$C$_{62}^{cage}$~+~C$_2$ & 9.4$^a$ / 9.0$^b$ & $2\times10^{20}$\\
\vspace{0.2cm}
{ (4)}&{C}$_{62}^{cage} \rightarrow~$C$_{60}^{cage}$~+~C$_2$ & 6.0$^a$ / 8.1$^b$ & $2\times10^{20}$\\
\vspace{0.2cm}
{ (5)}&{C}$_{60}^{cage} \rightarrow~$C$_{58}^{cage}$~+~C$_2$ & 11.1$^a$ / 11.2$^b$ & $2\times10^{20}$\\
\vspace{0.2cm}
{ (6)}&{C}$_{58}^{cage} \rightarrow~$C$_{56}^{cage}$~+~C$_2$ & 8.7$^a$ / 9.7$^b$ &$2\times10^{20}$\\
%+ 20 \text{H}
%\text{C}_{66}^{planar} \rightarrow \text{C}_{66}^{cage}
%\text{C}_{66}\rightarrow \text{C}_{64} + \text{C}_{2}
%\text{C}_{58}\rightarrow \text{C}_{56} + \text{C}_{2}
%C$_{60}\rightarrow$ C$_2$ + C$_{58}$ & $E_{60}=10.0$ & $k_{60}=10^{20.9}$\\
%PAH$_{n}\rightarrow$ C$_2$ + PAH$_{n-2}$ & $E_{PAH}=8.5$ & $k_{PAH}=10^{17.5}$\\
\hline
\end{tabular}
\tablefoot{{ a.} Using activation energies of \citet{zha92}; { b.} Using activation energies of \citet{glu04}.}
\end{center}
\end{table}

\begin{figure}
\begin{center}
\includegraphics[width=\hsize]{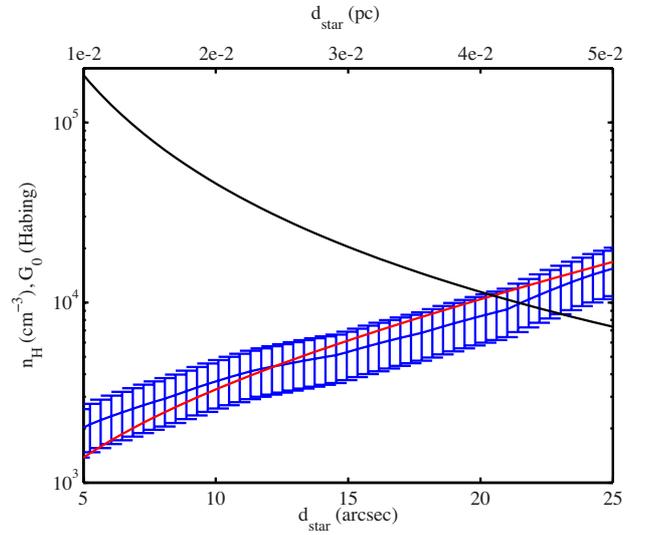}
\caption{ Physical conditions in NGC 7023. Density profile derived from observations of the far-infrared dust emission in the cavity of NGC 7023 (blue with error-bars) and exponential fit to this profile used in the photochemical model (red line). The black line shows the adopted profile for the intensity of the radiation field $G_0$.
 \label{fig_profile}}
\end{center}
\end{figure}

\begin{figure}
\begin{center}
\includegraphics[width=\hsize]{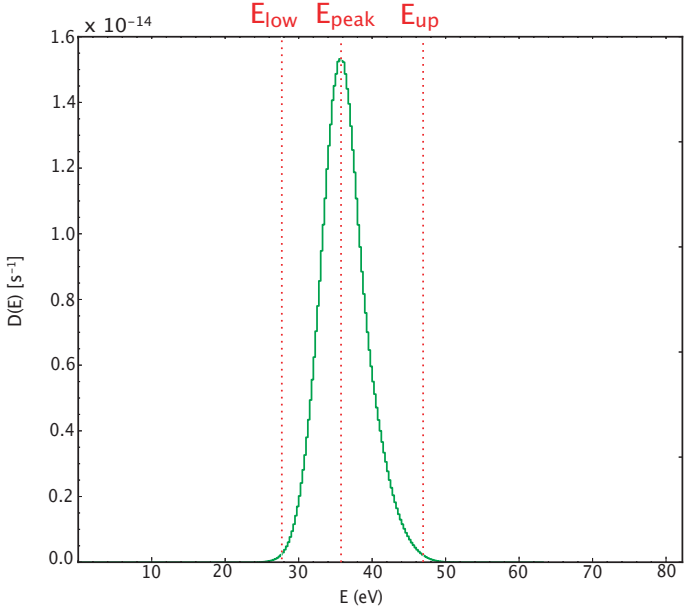}
\caption{Probability density function of dissociation of the \cs molecule as a function of internal energy, at a distance of 5'' from the star (G$_0\sim2\times10^5$). Energies $E_{peak}$ at maximum dissociation rate ($D(E)=k_{diss}(E)\times n(E)$), $E_{low}$ and $E_{up}$ (where $D(E_{up}) = D(E_{low}) = D(E_{peak})/100$ and $E_{up} > E_{low}$) are depicted. The values of $E_{peak}$, $E_{low}$ and $E_{up}$ for all cages are given in Table~\ref{tab_internal_energies}.
\label{fig_pdf}}
\end{center}
\end{figure}

\vspace{-0.5cm}

\begin{figure*}
\begin{center}
\includegraphics[width=14cm]{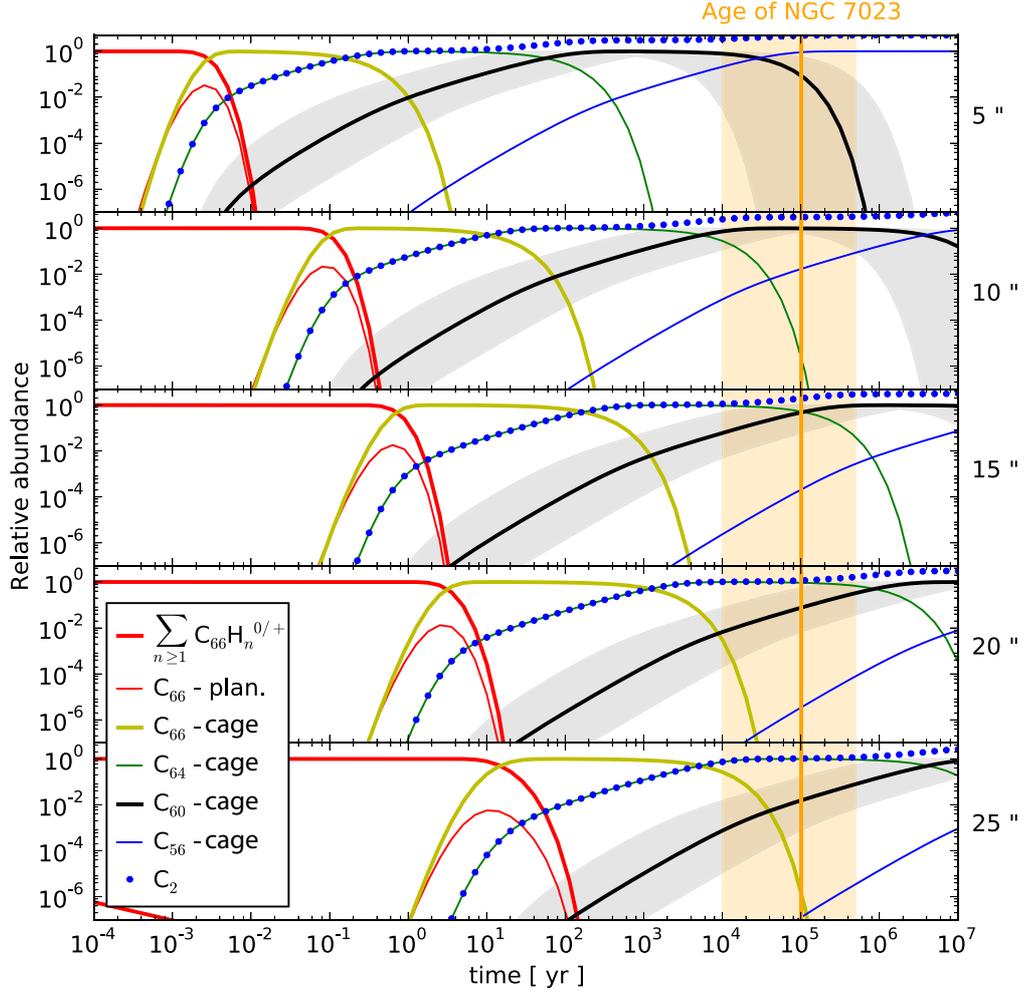}
\caption{Model results for the time-evolution of the abundance of
PAHs ($\sum\limits_{n=1}^{20}$C$_{66}$H$_{n}$), and cages ({C}$_{66}^{planar}$, {C}$_{66}^{cage}$, {C}$_{64}^{cage},${C}$_{62}^{cage}$, {C}$_{60}^{cage}$, 
{C}$_{58}^{cage}$) and C$_2$. These are normalized to the abundance of C$_{66}$H$_{20}$ at $t=0$. 
 The model calculations are done for distances of 5, 10, 15, 20 and 25'' from the star HD 200775 in NGC 7023. Note that the 
 dissociation of C$_2$ is not treated in our model, hence the C$_2$ abundances reported are only provided as an indication 
 of how much C$_2$ is formed from the dissociation of cages. Similarly, the dissociation of  the {C}$_{56}^{cage}$ is not included 
 in the model and therefore its abundance is provided as an indication of how much \cs is being destroyed. 
The gray shadowed regions indicate the uncertainty range implied by { uncertainties of a factor of 10 on the rates of C$_2$ loss} 
(see Sect.~\ref{sect_sensitivity} for details). The approximate age of NGC 7023 is given, with orange shadowed regions representing
the uncertainty on this value (see Sect.~\ref{sec:physical_conditions} for details). 
 \label{fig_res}}
\end{center}
\end{figure*}

%\subsection{Photochemical model}

\section{Astrophysical environment: NGC 7023}\label{sec:physical_conditions}

Since NGC 7023 is the template interstellar source for the study of fullerene formation \citep{sel10, ber12, mon13},
we will model the photochemical evolution of C$_{66}$H$_{20}$ for conditions found in this nebula. The formation of \cs 
is thought to occur inside the cavity between the star HD 200775 and the PDR situated 40'' at the North-West of the star.
Here, we study five positions at 5, 10, 15, 20 and 25'' from the star, situated on a cross-cut (see Fig. 1 in \citealt{mon13}) 
that goes from the star to the PDR. On the same cross-cut, we derive the spectral energy distribution of dust emission  
using archival data from the \emph{Herschel} space telescope (see \citealt{abe10}). Using these SEDs, we derive the column density on the line 
of sight using a value for dust opacity at 250 $\mu$m of 1.14$\times10^{-25}$cm$^{2}$ per H atom (see details on the 
method in \citealt{abe11}). This yields a column density profile, which we divide by the thickness of the region derived 
by \citet{pil12} based on the modeling of the PAH emission profile, i.e. 0.13 pc. The resulting density profile is shown 
in Fig.~\ref{fig_profile}. To obtain an analytical description of this profile, we fit it with an exponential law (see 
Fig.~\ref{fig_profile}). The intensities of the radiation field at 5, 10, 15, 20 and 25'' from the star are derived as in \citet{mon13} and the
corresponding profile of $G_0$ (UV field intensity in terms of the Habing field, which corresponds to $1.6 \times 10^{-3}$ 
erg cm$^{-2}$ s$^{-1}$ when integrated  between 91.2 and 240 nm, \citealt{hab68}) 
as a function of distance to the star is presented in Fig.~\ref{fig_profile}.
In order to derive the temperature of the gas at these positions PDR modeling is required. For the conditions described 
above the gas temperature derived by the Meudon PDR model typically ranges between a few 100 and 1000 K. Here we adopt a 
characteristic temperature of 300 K for the photochemical model but the exact value is not critical (see Sect.~\ref{sect_sensitivity}).
Finally, a parameter that is key when comparing the results of the model with observations is the age of the nebula, 
i.e. the time from when the cloud started to receive UV photons. This number cannot be derived directly and hence 
other indicators have to be used. One is the age of the illuminating star HD 200775. It is difficult to give a precise 
``age" for such a young and massive star, however, it is most likely ranging between $10^4$ and $5\times 10^5$ 
years (\citealt{ale08, ale13} and Alecian 2014 priv. comm.). \citet{ber12} adopted a value for the age of the nebula
ranging from a few $10^4$ to $1\times 10^5$ years. The analysis of the dynamical properties of the warm neutral
gas traced by the C$^{+}$ line with Herschel indicate an age of $\sim$ 0.5 Myrs \citep{ber12b}. However, 
more recent calculations focusing on the photo-evaporation flow in the NGC 7023 North PDR indicate ages as 
low as $1.6\times10^4$ years (Bern\'e et al. in prep). We will use $10^5$ years as the reference age, but we will
consider a range between $10^4$ and $5\times10^5$ as an age uncertainty range.

\section{Results}

\subsection{Internal energies}\label{internal_energies}

In order to understand the processes at play in the photochemistry, it is useful 
to evaluate the probability density functions (PDFs) of the internal energies of the species 
that photodissociate with the model. Fig.~\ref{fig_pdf}, shows such a PDF for \cs at a distance of 5'' from the
star. In Table~\ref{tab_internal_energies}, we report the main characteristics of these 
PDFs for all the cages considered in the model: the peak energy $E_{peak}$ i.e. the energy of maximum dissociation rate $D(E)$
expressed in s$^{-1}$, and the lower 
and upper bounds $E_{low}$ and $E_{up}$ defined by $D(E_{up}) = D(E_{low}) = D(E_{peak})/100$ with $E_{up} > E_{low}$
(see Fig.~\ref{fig_pdf} for a graphical definition of these parameters).
These values are calculated using the activation energies of \citet{zha92}, and those of \citet{glu04} (Table~\ref{tab_act}). 
The results presented in Table~\ref{tab_internal_energies} indicate that the internal energies required to dissociate the molecules 
are high, i.e. $>$ 15 eV. Photons in the NGC 7023 nebula have a maximum energy set by the Lyman limit, i.e. 13.6 eV.
Hence, the molecules that dissociate must have absorbed multiple photons, and therefore the
photochemistry of these species is completely governed by these multiple-photon absorptions.
In the specific case of \cs, the energy needed is at least 27 eV, i.e. requiring the 
absorption of at least three photons in order to dissociate.

\begin{table}

\caption{ Properties of the probability density functions of dissociation of cages (see example of such a function in Fig.~\ref{fig_pdf}) :
$E_{peak}$, $E_{low}$ and $E_{up}$. All values in eV and derived for a distance of 5'' from the star ($G_0\sim2\times10^5$).}
\label{tab_internal_energies}
\begin{center}
\begin{tabular}{lccccccc}
\hline \hline
& \multicolumn{3}{c}{{  a.}} & & \multicolumn{3}{c}{  b.} \\
\hline\hline
Species$^{*}$	& $E_{low}$	& $E_{peak}$	& $E_{up}$	& & $E_{low}$	&$E_{peak}$	& $E_{up}$\\
\hline
C$_{66}$    		&22       		&32     		&39       		& &26       		&35     		&43\\
C$_{64}$   		&24       		&35     		&43       		& &24       		&34     		&41\\
C$_{62}$   		&15       		&20     		&28       		& &22       		&30     		&36\\
C$_{60}$    		&27       		&36     		&47       		& &27       		&36     		&47\\
C$_{58}$    		&22       		&30    		&37       		& &24       		&33     		&41\\
\hline
\end{tabular}
\tablefoot{$^*$Cages only are presented
in this table. {  a.} Using activation energies of \citet{zha92}; {  b.} Using activation energies of \citet{glu04}. }
\end{center}
\end{table}

\subsection{Abundances}\label{abundances}

The results of the model are given in Fig.~\ref{fig_res} which presents the time-evolution of the abundance of
PAHs ($\sum\limits_{n=1}^{20}$C$_{66}$H$_{n}$), and cages ({C}$_{66}^{planar}$, {C}$_{66}^{cage}$, {C}$_{64}^{cage},${C}$_{62}^{cage}$, {C}$_{60}^{cage}$, 
{C}$_{58}^{cage}$) and C$_2$, at distances of 5, 10, 15, 20 and 25'' from the star HD 200775 in NGC 7023.
For these five positions, full dehydrogenation occurs very quickly (a few 10s of years at 25'' and a few days at 5'' from the 
star). This implies that C$_{66}$H$_{20}$ is quickly destroyed in NGC 7023, however larger PAHs could survive 
over longer timescales. Once the planar {C}$_{66}$ is formed, it immediately folds and forms a cage. This implies that
graphene flakes are only transient species and therefore unlikely to be detected in the ISM. The C$_{64}$ cage, instead,
can survive for a relatively long period of time (few 10s of years at 5'' from the star and up to several $10^5$ years at 25'' 
from the star). It is therefore the shrinking step, where cages loose C$_2$ units, that limits the efficiency of the 
\cs formation process in this model. {   Since each of the C$_2$ loss process is time consuming, it appears unlikely
that PAHs with sizes $N_C \gtrsim 66$ C atoms will contribute significantly to the formation of \cs.}
At distances shorter than 10'' from the star the cages can shrink to 
\cs and after $10^5$ years all the C$_{66}$H$_{20}$ has been converted to \cs. Once the molecules have reached C$_{60}$, 
it takes a very long time to destroy them: \cs survives for 10's of Myrs at distance larger than 10'' to the star
(radiation fields below $G_0$ of a few $10^4$). 
%This implies that after being formed, \cs can survive indefinitely
%in the diffuse ISM where $G_0\sim1$, unless other processes such as shocks or cosmic rays are able to dissociate it. 
Only at the closest position to the star (5'', $G_0 \sim 2\times10^5$) is \cs destroyed efficiently and hence it is
predicted that its abundance will decrease after a few $10^4$ years.

\subsection{Sensitivity of model results}\label{sect_sensitivity}

Is the model very sensitive to the adopted parameters, and if so, which are the critical parameters ?

First, we have checked that the dehydrogenation timescale is always much smaller than the cage
formation plus shrinking timescale for the physical conditions of the considered positions. Therefore,
it is equivalent to start our calculations with the planar C$_{66}$ rather than with C$_{66}$H$_{20}$.
All the following tests have been conducted with the above assumption{, except for the gas temperature test.
We first test the choice in the vibrational frequencies for the system. We find the exact values of the frequencies
to have a negligible impact on the results, as long as the number of vibrational degrees of freedom and the orders of
magnitude of the frequencies are correct. We have varied the gas temperature in the model from 100, to 300 and 
1000 K and found that this has a negligible impact on the \cs formation timescale. The UV-visible absorption cross 
section of the cages, which is not well known, has a somewhat larger impact on the 
results. For instance, if for the cages we use the cross section per C atom of C$_{66}$H$_{20}$ \citep{mal07} instead 
of the one of \cs \citep{ber99}, we obtain a variation by a factor of three in the timescale of formation of C$_{60}$. 
Multiplying the pre-exponential factors of C$_2$ loss by 0.1 or 10 simultaneously for all cages we observe an inversely 
proportional variation of the \cs formation timescale. This effect in the abundance of \cs is illustrated by the shadowed region
in Fig.~\ref{fig_res}. When varying individually these coefficients, we find that the results are impacted mainly by the value for C$_{66} \rightarrow$ 
C$_{64}$ + C$_2$, and marginally for C$_{64} \rightarrow$ C$_{62}$ + C$_2$, while changes for the two other shrinking 
reactions do not impact the results. We compared our standard results (i.e. using the activation energies from \citealt{zha92}) 
with the results obtained when  using the experimental dissociation energies of \citet{glu04}. The \cs formation timescales is slightly increased, mainly due
 to the higher stability of  {C}$_{66}^{cage}$ in the experimental data set. Overall, the effect does not affect the comparison
 with observations and discussion which follow. Hence from now on we only consider the results of the model using 
 the activation energies from \citealt{zha92}. The main source of uncertainty appears to arise from the pre-exponetial
 factors chosen for the shrinking steps. We will therefore consider their effect in the comparison with observations. 
}

\section{Comparison with observations}

%\subsection{Abundances}

In this section we compare the results of our model with the observation of \cs formation in 
NGC 7023. To be completely accurate, this comparison  would 
require that the model contains a whole size distribution of PAHs in agreement with observations. 
This requirement is beyond the scope of this paper, but the agreement between the observations and the model 
can be tested bearing this limitation in mind. 

The comparison is primarily based on confronting the maximal abundance of \cs observed in NGC 7023, at a distance of 11'' from the star,
with the value obtained in the model for a distance of 10''. Taking into account the uncertainty on the age of NGC 7023
and the uncertainty on the pre-exponential factors, the model-predicted abundances of \cs ranges between a 
minimum value of $1.3\times10^{-2}$ and 0.98, relative to C$_{66}$H$_{20}$ (Fig.~\ref{fig_res}). To convert this into 
an absolute abundance of \cs one needs to know the abundance of C$_{66}$H$_{20}$ relative to the total PAH population, 
and to know the total abundance PAHs in NGC 7023. The latter was estimated by \citet{ber12} to be 
$f_C^{PAH}=7\times10^{-2}$ expressed in fraction of carbon locked in PAHs. The maximum abundance of 
C$_{66}$H$_{20}$ relative to the total PAH population, $\alpha$, can be estimated following \citet{pil09},
using the PAH size distribution of \citet{des90}, which yields a value
$\alpha=5.3\times10^{-3}$.  Thus, the fraction of carbon locked in C$_{66}$H$_{20}$ is 
$5.3\times10^{-3}\times7\times10^{-2} = 3.7\times10^{-4}$. With this value, the fraction of carbon locked in \cs predicted by the model at a distance of
10'' from the star ranges between $f_C^{C_{60}}=4.8\times10^{-6}$ and $f_C^{C_{60}}=3.6\times10^{-4}$
(where the lower and upper limit of the range include the uncertainty on the pre-exponential factor and the age of the nebula). 
This range of values is in agreement with the one derived by \citet{ber12} of $f_C^{C_{60}}=1.7\times10^{-4}$.
So far we have considered  C$_{66}$H$_{20}$ as the only precursor of \cs. 
{   As mentioned in Sec.~\ref{abundances}, it is reasonable to expect that only PAHs bearing between 60 and  66 C-atoms will be able to 
form C$_{60}$, and hence can increase the final abundance of C$_{60}$.}
Hence, it is realistic to calculate $\alpha$ incorporating
all the species bearing between 60 and $\sim$66 C atoms, which yields $\alpha=3.8\times10^{-2}$. With this value, 
the abundance of \cs at 10'' from the star is predicted to range between $f_C^{C_{60}}=3.5\times10^{-5}$
and $f_C^{C_{60}}=2.6\times10^{-3}$, in good agreement with the  value derived by \citet{ber12} of 
$f_C^{C_{60}}=1.7\times10^{-4}$. The comparison can be extended to the other positions  (15, 20 and 25'') where 
the abundance of \cs has been measured. At these three positions, the model derived
ranges of abundance are also in good agreement with the observed abundances. However, as the distance grows
the range of acceptable values predicted by the model become broader and less constraining.
Overall, the present comparison demonstrates that our scenario is compatible with observations within the uncertainties 
on the molecular parameters and the age of the nebula. 

%Finally, \citet{cas14} recently conducted a study on \cs in a sample of PDRs. They find that \cs is detected in PDRs with intense UV fields, 
%$G_0 >1000$ (with the exception of Ced 201,see discussion in their paper) and its abundance ranges between 
%$f_C^{C_{60}}=4.2\times10^{-5}$ and $f_C^{C_{60}}=5.6\times10^{-4}$. Although a detailed comparison cannot be done since our 
%model was developed specifically for NGC 7023, these ranges of abundances appear in good agreement with those discussed hereabove. 

%A particularly interesting aspect of our results is that we are able to reproduce the high 
%abundances of \cs found at the positions close to the star.

%\begin{figure}
%\begin{center}
%\includegraphics[width=\hsize]{Comp_obs_model.eps}
%\caption{ Comparison between the observed efficiency of conversion of C$_{66}$H$_{20}$ to \cs $\epsilon_{obs}$ (normalized so 
%that $\epsilon_{obs}(10'')=1$) and the efficiency derived from the photochemical model  $\epsilon_{mod}$ (in red) given as a function
%of distance to the star. 
%\label{fig_comp_obs}}
%\end{center}
%\end{figure}

\section{Discussion}\label{discussion}

\subsection{Comparison to other models}
Models of \cs formation in a top-down mechanism were presented in
\citet{ber12} and \citet{mic12}.  The scenarios in these papers are quite similar, the main difference 
being that \citet{ber12} consider PAHs as the starting ingredient while for \citet{mic12} it is nanometer-sized ``arophatic" clusters. 
In both of these studies the evolution of the species is  described by a ``thermal model" (Arrhenius equation),
where the driving parameter is the activation energy. For the loss of C atoms at the edge of graphene sheets, 
\citet{ber12} used a value tuned to 4.5 eV to obtain reasonable formation efficiencies. Yet, as noted by \citet{mic12} 
this value is somewhat arbitrary and the obtained \cs formation efficiencies remain low.
To explain the efficient formation of \cs in evolved stars, \citet{mic12} used the results from the molecular 
dynamics simulations of \citet{zhe07}, and extracted a unique activation energy of 0.35 eV
for the shrinking reaction by loss of C$_2$.  However, this value is a factor 
of $\sim20$ smaller than what is generally accepted for this reaction (see Table 1), leading to over-estimations 
of the rates by many orders of magnitude, casting serious doubts on the conclusions of \citet{mic12}. 
In summary, since these simple models do not allow
the molecules to reach high internal energies through multiple photon absorptions, high formation 
rates of \cs require activation energies tuned to values that are too low to be physical.

In summary, since these simple models do not consider multiple photon absorptions, the molecules never reach 
the high internal energies necessary to a high formation rate of \cs. Therefore, 
they strongly underestimate the \cs formation yields, unless one assumes activation energies that are too low to 
be physical. Instead, the approach presented here,
including a complete description of the photochemical processes, allows, using realistic molecular parameters, 
to predict formation efficiencies that are in agreement with observations.

\subsection{Comparison to experimental results of gas-phase cage formation}

{    After we submitted the present paper, \citet{zhe14} reported experimental results in which they demonstrate 
that \cs can be formed in the gas phase through laser irradiation of larger PAH molecules. These results support 
the idea put forward here that \cs can be formed following a top-down scheme. However, \citet{zhe14}  suggest that 
in some cases the conversion of graphene into cages could involve a prior step of C$_2$ loss. This is in contradiction 
with our photochemical model in which the dissociation by loss of C$_2$ is less efficient than the folding by 
many orders of magnitude, and therefore folding always precedes the loss of C$_2$. This is mainly because
the activation energy for folding  (which relies on molecular parameters derived from molecular dynamics simulations 
performed by \citealt{leb12}) is a factor of $\sim$ 2 smaller than the activation energy for C$_2$ loss. 
Further experimental investigation is necessary, in particular to quantify the activation energies involved in the
folding and C$_2$ loss by flakes, so that they can be included in our model.}

%of obtaining the complete probability density function of the internal energies of the molecules
%relying on both UV absorption and relaxation 
\subsection{Stability of \cs}

Table~\ref{tab_internal_energies} demonstrates that \cs can reach particularly high internal energies. In this state, fluorescence
from thermally excited electronic states becomes a very efficient cooling channel, postponing the dissociation and conferring an increased stability to
\cs \footnote{Note that since multiple photon absorptions are rare, most \cs molecules in the ISM are situated
at much lower internal energies and the population-averaged  emission at a given time remains dominated by infrared
transitions. The visible emission of \cs in the ISM is therefore probably very weak.}. Hence, in the frame of the top-down mechanism 
detailed here and because of its remarkable stability, \cs is expected to be the most abundant of all fullerenes in highly UV irradiated environments in space.
It is interesting to note that similar kinetic stability arguments have been put forward \citep{fed11} to explain the high abundance of \cs relative to 
other fullerenes observed in laboratory experiments \citep{kro85, kra90}. 
As discussed in Sect. \ref{abundances}, \cs is expected to survive for 10s of Myrs in intense radiation fields. 
In the diffuse ISM, where the radiation field is several orders of magnitude smaller, \cs is therefore expected to resist to radiation for even
larger timescales. In such conditions, other energetic processes must be invoked to destroy \cs, such as shocks or 
cosmic rays.

\subsection{Isomerization of \cs}

There exists a large number of \cs isomers, and hence there is {\it a-priori} no 
reason for the icosahedral $Ih$-\cs to be the one present in the ISM. $Ih$-\cs has the lowest energy amongst all \cs isomers 
and the closest isomers lie about 2 eV higher in energy \citep{rag92}. Upon UV absorption, the isomerization reaction 
from any \cs isomer towards $Ih$-\cs  requires only 5.4 eV of activation energy  \citep{yi92}, and is therefore expected be fast 
in the conditions investigated here. Conversely, the isomerization reaction from $Ih$-\cs to another \cs isomer requires 
$\sim$ 5.4 + 2 = 7.4 eV, and will be much slower than the previous reaction. Therefore, relatively quickly, the population of \cs 
molecules will be dominated by the icosahedral isomer.

\section{Conclusions}

We have presented the first detailed photochemical modeling of \cs formation from PAHs in space following a top down 
scheme. This model is calculated for a single molecule (C$_{66}$H$_{20}$) as a starting point, and the key 
processes are explicitly described: UV photon absorption (including multiple photon events), radiative cooling and dissociation. 
Because the involved activation energies in this top-down 
mechanism are high (especially for the shrinking steps), multiple photon absorptions dominate the photochemistry. 
Using the physical conditions which prevail in the NGC 7023 
reflection nebula, we find that \cs can be formed from C$_{66}$H$_{20}$ over a timescale of about $10^5$ years
with a high efficiency in highly irradiated regions.  {  Assuming that only PAHs containing between 60 and 66 C atoms
are precursors of \cs and} a classical size distribution for PAHs, the comparison between the modeled
and observed abundances of \cs shows a good agreement, within the uncertainties of the model. 
These uncertainties can be reduced once a better characterization of the reaction rates of the shrinking of 
cages is available. While developed for the physical conditions of NGC 7023 as representative 
of the interstellar medium, the scenario and model described in this paper could also be applied to 
the irradiated circumstellar gas of planetary nebulae, if the physical conditions in these environments can be 
characterized in detail.

%\vspace{-0.5cm}

\begin{acknowledgements}
This work was supported by the CNRS program ``Physique et Chimie du Milieu Interstellaire" (PCMI)
JM acknowledges the support of the Academy of Finland grant No. 250741, and the support of the University of Franche-ComtŽ\'e through the BQR funding.
We also acknowledge support from the European Research Council under the European 
Union's Seventh Framework Programme ERC-2013-SyG, Grant Agreement n. 610256 NANOCOSMOS.
We acknowledge the referee for constructive comments which improved the quality of this manuscript.
\end{acknowledgements}

\bibliographystyle{aa}
\bibliography{biblio}

%\begin{appendix} %First online appendix

%\end{appendix}

\end{document}